\documentclass[dvipdfmx,aps,prd,notitlepage,nofootinbib,preprintnumbers,superscriptaddress]{revtex4}
\usepackage{amssymb}
\usepackage{amsmath}
\usepackage{verbatim}
\usepackage{appendix}
\usepackage{hyperref}

\begin{document}
\preprint{IPMU18-0208, YITP-18-131}
\title{Kinetic equation for Lifshitz scalar}

\author{Shinji Mukohyama}
\affiliation{Center for Gravitational Physics, Yukawa Institute for Theoretical Physics, Kyoto University, 606-8502, Kyoto, Japan}
\affiliation{Kavli Institute for the Physics and Mathematics of the Universe (WPI), The University of Tokyo Institutes for Advanced Study, The University of Tokyo, Kashiwa, Chiba 277-8583, Japan}
\affiliation{Institut Denis Poisson, UMR - CNRS 7013, Universit\'{e} de Tours, Parc de Grandmont, 37200 Tours, France}

\author{Yota Watanabe}
\affiliation{Kavli Institute for the Physics and Mathematics of the Universe (WPI), The University of Tokyo Institutes for Advanced Study, The University of Tokyo, Kashiwa, Chiba 277-8583, Japan}

\date{\today}

\begin{abstract}
 Employing the method of Wigner functions on curved spaces, we study classical kinetic (Boltzmann-like) equations of distribution functions for a real scalar field with the Lifshitz scaling. In particular, we derive the kinetic equation for $z=2$ on general curved spaces and for $z=3$ on spatially flat spaces under the projectability condition $N=N(t)$, where $z$ is the dynamical critical exponent and $N$ is the lapse function. We then conjecture a form of the kinetic equation for a real scalar field with a general dispersion relation in general curved geometries satisfying the projectability condition, in which all the information about the nontrivial dispersion relation is included in the group velocity and which correctly reproduces the equations for the $z=2$ and $z=3$ cases as well as the relativistic case. The method and equations developed in the present paper are expected to be useful for developments of cosmology in the context of Ho\v{r}ava-Lifshitz gravity. 
\end{abstract}

\maketitle


\section{Introduction}
\label{sec:intro}

Gravitational phenomena observed to the present are well described by general relativity (GR). However, at short distances quantum gravity effects become significant and GR loses its predictability. Among various approaches to quantum gravity, Ho\v{r}ava-Lifshitz (HL) gravity is unique in the sense that it is a local field theory of gravity which is perturbatively renormalizable at least at the power-counting level and which is free from the Ostrogradsky ghost \cite{Horava:2009uw}. The theory is rendered power-counting renormalizable by the scaling anisotropic in space and time (the so-called Lifshitz scaling) at high energy,
\begin{equation}
 t \to b^z t\,, \quad \vec{x} \to b \vec{x}\,,
\end{equation}
and recovers the usual isotropic scaling at low energy. Here, $z$ is a constant often called the dynamical critical exponent and $b$ is an arbitrary constant. The renormalizability beyond the power-counting level has recently been proven in the minimal setup called the projectable theory, where the lapse function depends only on time \cite{Barvinsky:2015kil,Barvinsky:2017zlx}. As for observational constraints on parameters, see, e.g., Refs.~\cite{Gumrukcuoglu:2017ijh,Ramos:2018oku} for the nonprojectable theory and Sec.~3.2 of Ref.~\cite{Mukohyama:2010xz} for the projectable theory.

Since HL gravity is a candidate for quantum gravity and thus can presumably describe gravity at very short distances, it is natural to seek its implications to the early universe cosmology~\cite{Mukohyama:2010xz}. In the early universe or at high energies, it is expected from theoretical consistency that not only the gravity sector but also the matter sector should exhibit the Lifshitz scaling. A field in the matter sector should then obey a dispersion relation of the form $\omega^2 \sim (k^2/a^2)^z/M^{2(z-1)}$ in the early universe, where $k$ is the comoving momentum, $a$ is the scale factor of the universe, $M$ is the (momentum) scale characterizing the Lifshitz scaling, and $z$ is the dynamical critical exponent. The energy of each high energy particle should decay as $\omega \propto 1/a^{z}$ as the universe expands. As a result the energy density of a gas of Lifshitz particles should decay as $\rho_z \propto 1/a^{z+3}$~\cite{Mukohyama:2009zs}. In reality, we need to consider the transition from a high value of $z$ to lower values, say, from $z=3$ to $z=2$ and then to $z=1$ in the course of cosmic expansion by taking into account the dispersion relation of the form $\omega^2 = \alpha (k^2/a^2)^3/M^4 + \beta (k^2/a^2)^2/M^2 + \gamma (k^2/a^2) + m^2$, where $\alpha$, $\beta$, and $\gamma$ are constants of order unity and $m$ is the field's mass. It is naturally expected that terms with higher $z$ should be important/unimportant at higher/lower temperature. Thus, it is expected that the $z=3$ behavior is realized at a temperature higher than $M$ and then the system exhibits $z=2$ and $z=1$ behaviors at lower temperature. A more quantitative analysis of the behavior of the system requires a kinetic equation for the distribution function in the phase space, namely a generalization of the Boltzmann equation to a gas of particles with the Lifshitz scaling at high energies. Moreover, considering the fact that the Boltzmann equation has been one of the essential tools for the analysis of cosmological perturbations in GR, it is obvious that the generalization of the Boltzmann equation will be useful for further developments of cosmology in the context of Ho\v{r}ava-Lifshitz gravity.

The aim of the present paper is, as the first step, to derive a kinetic equation for a matter field with the Lifshitz scaling in a fixed background spacetime. For simplicity we shall consider especially a real scalar field. The first-principle derivation of a kinetic equation in Lorentz-invariant theories was developed in Minkowski spacetime \cite{Groot:1980} by using the Wigner function, which is the Fourier transformed two-point function with respect to the relative distance between the two points \cite{Balazs:1983hk,Hillery:1983ms}. While its generalization to curved spacetimes has been studied \cite{Winter:1986da,Calzetta:1987bw,Fonarev:1993ht,Antonsen:1997dc,Prokopec:2017ldn,Friedrich:2018qjv}, the treatment of Ref.~\cite{Friedrich:2018qjv} is particularly suitable for our aim since it adopts a spatially covariant formalism based on the 3+1 decomposition of the spacetime and considers a real scalar field, which is of our interest in the present paper.

The rest of the present paper is organized as follows. In Sec.~\ref{sec:review} we review the derivation of a kinetic equation for a Lorentz-invariant real scalar field. We apply the method to derive a kinetic equation for a real Lifshitz scalar in Sec.~\ref{sec:Lifshitz scalar}. Section~\ref{sec:summary} is devoted to a summary and discussions. In Appendix \ref{sec:smallness} we argue that the subdominant Wigner functions remain small if they are set to initially. Energy, momentum, and stress tensors for Lifshitz scalar are given in Appendix \ref{sec:energy-momentm}.

\section{Derivation of relativistic Boltzmann equation}
\label{sec:review}

In this section we review the derivation of the kinetic (collisionless Boltzmann or Vlasov) equation for a free real scalar field $\phi$ at leading order in $\hbar$, based on Ref.~\cite{Friedrich:2018qjv}. Let us introduce Wigner functions,
\begin{equation}
F_{XY}(x^\mu,p_i):=\sqrt{\gamma}\int d^3r\, e^{-\frac{i}{\hbar}r^ip_i}\left\langle:\left[\exp\left(\frac{r^i}{2}\tilde{\nabla}_i\right)X(x^\mu)\right]\left[\exp\left(-\frac{r^i}{2}\tilde{\nabla}_i\right)Y(x^\mu)\right]:\right\rangle,
\end{equation}
where $x^\mu=(t,x^i)$, $t$ is the time coordinate, $x^i$ are spatial coordinates, $X, Y\in\{\phi,\phi_\perp\}$, $\phi_\perp:=\partial_\perp\phi$, $\partial_\perp=N^{-1}(\partial_t-N^i\partial_i)$,  $N$ is the lapse function, $N^i$ is the shift vector, $\gamma$ is the determinant of the spatial metric $\gamma_{ij}$, $:\cdots:$ denotes the normal ordering, $\langle\ldots\rangle$ denotes the expectation value,
\begin{equation}
\tilde{\nabla}_i:=\nabla_i-\Gamma_{ij}^k r^j \frac{\partial}{\partial r^k},\footnote{$\tilde{\nabla}_i$ in this article represents $\nabla^H_i$ in Ref.~\cite{Friedrich:2018qjv}.}
\end{equation}
$\nabla_i$ is the covariant derivative compatible with $\gamma_{ij}$, and $\Gamma_{ij}^k$ is the Christoffel symbols composed of $\gamma_{ij}$. Note that $\tilde{\nabla}_i r^j=0$ and
\begin{equation}
e^{\frac{r^i}{2}\tilde{\nabla}_i}X(x^\mu)=\left(1+\frac{r^i}{2}\nabla_i+\frac{r^ir^j}{8}\nabla_i\nabla_j+\cdots\right)X(x^\mu),
\end{equation}
which is a scalar function $X(x^\mu)$ acted on by a spatially covariant shift operator. 

Let us take linear combinations of the Wigner functions to make them real and have the same dimension\footnote{The factor $1/(2\pi\hbar)^3$ is not included contrary to Ref.~\cite{Friedrich:2018qjv}}:
\begin{align}
\label{eq:f_1^p}
 f_1^+:=&\frac{1}{2\hbar}\left[\frac{\omega}{\hbar}F_{\phi\phi} + \frac{\hbar}{\omega}F_{\phi_\perp\phi_\perp}\right], \\
 \label{eq:f_1^m}
f_1^-:=&\frac{i}{2\hbar}\left[F_{\phi_\perp\phi} - F_{\phi\phi_\perp}\right], \\
f_2:=&\frac{1}{2\hbar}\left[\frac{\omega}{\hbar}F_{\phi\phi} - \frac{\hbar}{\omega}F_{\phi_\perp\phi_\perp}\right], \\
\label{eq:f_3}
f_3:=&\frac{1}{2\hbar}\left[F_{\phi_\perp\phi} + F_{\phi\phi_\perp}\right],
\end{align}
where $\omega$ is the energy, which is in the Lorentz-invariant massless case $\omega=\sqrt{\gamma^{ij}p_ip_j}$.

One can see that $f_1:=f_1^++f_1^-$ corresponds to the classical distribution function for particles and $f_2$, $f_3$ express quantum and field-theoretical corrections from the following discussion. For a free real scalar described by the diffeomorphism invariant action 
\begin{equation}
I=\int d^4x\, \sqrt{-g} \left[-\frac{1}{2}g^{\mu\nu}\partial_\mu \phi \partial_\nu\phi \right],
\end{equation}
the energy-momentum tensor
\begin{equation} \label{eq:Tmunu}
T^{\mu\nu}=\frac{2}{\sqrt{-g}}\frac{\delta I}{\delta g_{\mu\nu}}
\end{equation}
is expressed with the $3+1$ decomposition as
\begin{equation}
T_{\mu\nu}=E n_\mu n_\nu +P_\mu n_\nu+P_\nu n_\mu +S_{\mu\nu}, 
\end{equation}
where $n_\mu=(-N,0)$ and\footnote{It should be understood that $d^3p$ represents $\prod_{i=1}^{3}dp_i$.} \cite{Friedrich:2018qjv} 
\begin{align}
\label{eq:E}
\left\langle:E:\right\rangle=&\int \frac{d^3p}{(2\pi\hbar)^3\sqrt{\gamma}}\left[\omega f_1+\mathcal{O}(\hbar)\right], \\
\left\langle:P_i:\right\rangle=&\int \frac{d^3p}{(2\pi\hbar)^3\sqrt{\gamma}}\left[p_i f_1+\mathcal{O}(\hbar)\right], \\
\label{eq:Sij}
\left\langle:S_{ij}:\right\rangle=&\int \frac{d^3p}{(2\pi\hbar)^3\sqrt{\gamma}}\left[\frac{p_ip_j}{\omega}f_1+\mathcal{O}(f_2)+\mathcal{O}(\hbar)\right]\,.
\end{align}
These expressions show that $E$ and $P_i$ depend only on $f_1$ at the leading order, and so does $S_{ij}$ if $|f_2|\ll |f_1|$, which will be justified in Appendix~\ref{sec:smallness}. (See Appendix~\ref{sec:energy-momentm} for generalization of the formulas (\ref{eq:E})--(\ref{eq:Sij}) to a general dispersion relation.) We also impose that $f_3$ is small when we derive the Boltzmann equation later. Therefore, the Boltzmann equation should be derived by computing the time derivative of $f_1$. On the other hand, the number density operator can be obtained by observing
\begin{equation}
\left\langle:\phi^2(x^\mu):\right\rangle=\int \frac{d^3p}{(2\pi\hbar)^3\sqrt{\gamma}}F_{\phi\phi}.
\end{equation}

To find derivative operators acting on the Wigner function $f_1$, we use the following trick. Define $X_E$ and $X_R$ for a given $3$-vector $X^i(x^\mu, r^j)\partial_i$ as
\begin{align}
X_E:=&X^i e_i, \\
X_R:=&X^i \frac{\partial}{\partial r^i},
\end{align}
where
\begin{equation}
e_i=\partial_i-\Gamma_{ij}^k r^j \frac{\partial}{\partial r^k},
\end{equation}
and also
\begin{equation}
f^\pm:=\exp\left(\pm\frac{r^i}{2}\tilde{\nabla}_i\right)f(x^\mu),
\end{equation}
for any scalar function $f(x^\mu)$ on spacetime. We have 
\begin{equation}
\label{eq:P^pmf^mp}
\mathcal{P}^\pm[X] f^\mp=0,
\end{equation}
where
\begin{equation}
\label{eq:P^pm}
\mathcal{P}^\pm[X]:=\pm X_R + \frac{1}{2}X_E - \frac{1}{2}\tilde{\nabla}_{r_E}X_R \pm \sum_{n=2}^{\infty}H_n^\mp[X],
\end{equation}
\begin{equation}
H_n^\pm[X]=\frac{1}{n! (\pm2)^n}[r^{i_1}e_{i_1},[r^{i_2}e_{i_2},[\cdots,[r^{i_n}e_{i_n},X_R]\cdots]]],
\end{equation}
which can be shown by applying the Baker-Campbell-Hausdorff formula to $X_R f^{\mp}=[X_R,e^{\mp\frac{r^i}{2}\tilde{\nabla}_i}]f$. Here, $r_E:=r^ie_i$, $\tilde{\nabla}_{r_E}:=r^i\tilde{\nabla}_i$, and $\tilde{\nabla}_{r_E}X_R$ denotes $(\tilde{\nabla}_{r_E}X)_R=r^i(\tilde{\nabla}_i X^j)\frac{\partial}{\partial r^j}$. It follows that
\begin{align}
X_E=&\mathcal{P}^+[X]+\mathcal{P}^-[X]+\tilde{\nabla}_{r_E}X_R+2\sum_{n=1}^{\infty}H_{2n+1}^+[X], \\
\label{eq:X_R}
X_R=&\frac{1}{2}\left(\mathcal{P}^+[X]-\mathcal{P}^-[X]\right)-\sum_{n=1}^{\infty}H_{2n}^+[X].
\end{align}
We see below that terms higher order in $r^i$ are higher order in $\hbar$ and thus are negligible in the derivation of the classical kinetic equation.\footnote{In other words, we take an expansion where $x^i$ derivatives on Wigner functions are negligible compared to $r^i$ derivatives on them.} Using \eqref{eq:P^pm} and \eqref{eq:X_R}, we find
\begin{align}
\tilde{\nabla}_{r_E}X_R&=\tilde{\nabla}_{r_E}\left[\frac{1}{2}\left(\mathcal{P}^+[X]-\mathcal{P}^-[X]\right)+\mathcal{O}(r)\right] \nonumber \\
&=\frac{1}{2}\left( \mathcal{P}^+[\tilde{\nabla}_{r_E}X] - \mathcal{P}^-[\tilde{\nabla}_{r_E}X] \right) +\mathcal{O}(r^2).
\end{align}
Define $X_{E/R}^\pm$ as $\mathcal{P}^\pm$ parts of $X_{E/R}$,
\begin{align}
X_E^\pm:=&\mathcal{P}^\pm[X] \pm \frac{1}{2}\mathcal{P}^\pm[\tilde{\nabla}_{r_E}X]+\mathcal{O}(r^2) \\
X_R^\pm:=&\pm\frac{1}{2}\mathcal{P}^\pm[X] +\mathcal{O}(r)
\end{align}
Since $X_E=X_E^++X_E^-$, we have
\begin{align}
\label{eq:X_Ep}
(X_Ef^+)g^-=&X_E^+(f^+g^-), \\
\label{eq:X_Em}
f^+(X_Eg^-)=&X_E^-(f^+g^-),
\end{align}
and similar identities hold for $X_R$. Then one can reexpress $X_{E/R}^\pm$ in terms of $X_E$ and $X_R$ as
\begin{align}
\label{eq:X_Epm}
X_E^\pm=&\pm X_R+\frac{1}{2}X_E+\mathcal{O}(r), \\
\label{eq:X_Rpm}
X_R^\pm=&\frac{1}{2}X_R+\mathcal{O}(r^0).
\end{align}

Now let us calculate time derivatives of the Wigner functions.
\begin{equation}
\label{eq:del_tF}
\partial_t F_{XY}=\sqrt{\gamma}\partial_t(F_{XY}/\sqrt{\gamma})+(NK+\nabla_iN^i)F_{XY},
\end{equation}
where $K=\gamma^{ij}K_{ij}$, $K_{ij}=(\dot{\gamma}_{ij}-\nabla_i N_j-\nabla_j N_i)/(2N)$\footnote{Note that $K_{ij}$ has the opposite sign from Ref.~\cite{Friedrich:2018qjv}.}. In order to compute the first term in the right-hand side of (\ref{eq:del_tF}), one needs to evaluate the time derivatives of $\phi^\pm:=\exp(\pm r^i\tilde{\nabla}_i/2)\phi$ and $\phi_\perp^\pm:=\exp(\pm r^i\tilde{\nabla}_i/2)\phi_\perp$. For this purpose, one can use the following formula for a general function $f$: 
\begin{align}
\partial_t  f^\pm&=[\partial_t,e^{\pm\frac{r^i}{2}\tilde{\nabla}_i}] f + e^{\pm\frac{r^i}{2}\tilde{\nabla}_i} (N f_\perp+N^i\partial_i f) \nonumber \\
&=[\partial_t,e^{\pm\frac{r^i}{2}\tilde{\nabla}_i}] f + N^ie_i f^{\pm} - [N^ie_i,e^{\pm\frac{r^i}{2}\tilde{\nabla}_i}] f + e^{\pm\frac{r^i}{2}\tilde{\nabla}_i}(N f_\perp) \nonumber \\
&=\mathcal{T}_\pm f^\pm+\mathcal{M}_\pm f^\pm+N^\pm f_\perp^\pm,
\end{align}
where $\mathcal{M}_\pm=N^ie_i-\tilde{\mathcal{M}}_\pm$,
\begin{align}
\mathcal{T}_\pm&=-\sum_{n=1}^\infty\frac{1}{n! (\pm2)^n}[r^{i_1}e_{i_1},[r^{i_2}e_{i_2},[\cdots,[r^{i_n}e_{i_n},\partial_t]\cdots]]], \\
\tilde{\mathcal{M}}_\pm&=-\sum_{n=1}^\infty\frac{1}{n! (\pm2)^n}[r^{i_1}e_{i_1},[r^{i_2}e_{i_2},[\cdots,[r^{i_n}e_{i_n},N^ie_i]\cdots]]],
\end{align}
and $e^{\pm\frac{r^i}{2}\tilde{\nabla}_i}(N f_\perp)=N^\pm f_\perp^\pm$ is used.

Defining $\mathcal{T}_\pm^\pm$ and $\mathcal{M}_\pm^\pm$ by expressing $\mathcal{T}_\pm$ and $\mathcal{M}_\pm$ as linear combinations of $\{ e_i, \partial/\partial r^j\}$ and then applying (\ref{eq:X_Epm}) or (\ref{eq:X_Rpm}) to each term, we have, similar to \eqref{eq:X_Ep} and \eqref{eq:X_Em},
\begin{align}
(\mathcal{M}_+f^+)g^-=&\mathcal{M}_+^+(f^+g^-), \\
f^+(\mathcal{M}_-g^-)=&\mathcal{M}_-^-(f^+g^-).
\end{align}
Similar identities hold for $\mathcal{T}_\pm^\pm$. Then
\begin{align}
\sqrt{\gamma}\partial_t(F_{\phi\phi}/\sqrt{\gamma})=&\frac{1}{2}\sqrt{\gamma}\int d^3r\, e^{-\frac{i}{\hbar}r^ip_i} (N^++N^-)\left\langle:\phi_\perp^+\phi^- + \phi^+\phi_\perp^-:\right\rangle \nonumber \\
&+\frac{1}{2}\sqrt{\gamma}\int d^3r\, e^{-\frac{i}{\hbar}r^ip_i} (N^+-N^-)\left\langle:\phi_\perp^+\phi^- - \phi^+\phi_\perp^-:\right\rangle \nonumber \\
 &+\sqrt{\gamma}\int d^3r\,e^{-\frac{i}{\hbar}r^ip_i}(\mathcal{T}_+^+ + \mathcal{T}_-^- + \mathcal{M}_+^+ + \mathcal{M}_-^-)\left\langle:\phi^+\phi^-:\right\rangle\,.
 \label{eqn:dFphiphi}
\end{align}
Hereafter we focus on the case with space-independent lapse $N=N(t)$, which implies that $N^+ = N^- = N$. Using the equation of motion,
\begin{equation}\label{eqn:eom_relativistic}
\partial_\perp^2\phi+K\partial_\perp\phi-\Delta\phi =0,
\end{equation}
where $\Delta=\gamma^{ij}\nabla_i\nabla_j$, we have
\begin{align}
\sqrt{\gamma}\partial_t(F_{\phi_\perp\phi}/\sqrt{\gamma})=&\sqrt{\gamma}\int d^3r\, e^{-\frac{i}{\hbar}r^ip_i} \left\langle:[(\mathcal{T}_+ + \mathcal{M}_+ -(NK)^+)\phi_\perp^+ + (N\Delta)^+\phi^+]\phi^- + \phi_\perp^+[(\mathcal{T}_- + \mathcal{M}_-)\phi^- + (N\phi_\perp)^-]:\right\rangle,
\end{align}
\begin{align}
\frac{1}{2}\sqrt{\gamma}\partial_t((F_{\phi_\perp\phi}-F_{\phi\phi_\perp})/\sqrt{\gamma})=&\frac{1}{2}\sqrt{\gamma}\int d^3r\, e^{-\frac{i}{\hbar}r^ip_i} (\mathcal{T}_+^+ + \mathcal{T}_-^- + \mathcal{M}_+^+ + \mathcal{M}_-^--\frac{1}{2}(NK)^+-\frac{1}{2}(NK)^-)\left\langle:\phi_\perp^+\phi^- - \phi^+\phi_\perp^-:\right\rangle \nonumber \\
&-\frac{1}{2}\sqrt{\gamma}\int d^3r\, e^{-\frac{i}{\hbar}r^ip_i} (\frac{1}{2}(NK)^+-\frac{1}{2}(NK)^-)\left\langle:\phi_\perp^+\phi^- + \phi^+\phi_\perp^-:\right\rangle \nonumber \\
&+\frac{1}{2}\sqrt{\gamma}\int d^3r\,e^{-\frac{i}{\hbar}r^ip_i}((N\Delta)_*^+-(N\Delta)_*^-)\left\langle:\phi^+\phi^-:\right\rangle,
\end{align}
where $\Delta_*^\pm$ are defined as $\Delta_*^+(f^+g^-)=(\Delta f)^+g^-$, $\Delta_*^-(f^+g^-)=f^+(\Delta g)^-$. Also,
\begin{align}
\sqrt{\gamma}\partial_t((F_{\phi_\perp\phi_\perp}/\sqrt{\gamma})=&\sqrt{\gamma}\int d^3r\, e^{-\frac{i}{\hbar}r^ip_i} (\mathcal{T}_+^+ + \mathcal{T}_-^- + \mathcal{M}_+^+ + \mathcal{M}_-^--(NK)^+-(NK)^-)\left\langle:\phi_\perp^+\phi_\perp^-:\right\rangle \nonumber \\
&-\frac{1}{2}\sqrt{\gamma}\int d^3r\,e^{-\frac{i}{\hbar}r^ip_i}((N\Delta)_*^+-(N\Delta)_*^-)\left\langle:\phi_\perp^+\phi^- - \phi^+\phi_\perp^-:\right\rangle \nonumber \\
&+\frac{1}{2}\sqrt{\gamma}\int d^3r\,e^{-\frac{i}{\hbar}r^ip_i}((N\Delta)_*^++(N\Delta)_*^- )\left\langle:\phi_\perp^+\phi^- + \phi^+\phi_\perp^-:\right\rangle,
\end{align}
At the leading order of expansion in $r^i$,
\begin{align}
\tilde{\mathcal{M}}_\pm&=\mp\frac{1}{2}[r_E, N^ie_i]+\mathcal{O}(r^2) \nonumber \\
&=\mp\frac{1}{2}\left(r^i(\tilde{\nabla}_iN^j)e_j + R^l_{\,ikj}r^ir^jN^k\frac{\partial}{\partial r^l}\right)+\mathcal{O}(r^2)
\end{align}
$\mathcal{M}_\pm=N^ie_i-\tilde{\mathcal{M}}_\pm$,
\begin{align}
\mathcal{M}_\pm&=N^ie_i\pm\frac{1}{2}\left(r^i(\tilde{\nabla}_iN^j)e_j + R^l_{\,ikj}r^ir^jN^k\frac{\partial}{\partial r^l}\right)+\mathcal{O}(r^2)
\end{align}
Putting $\pm$ and using \eqref{eq:X_Epm} and \eqref{eq:X_Rpm},
\begin{align}
\mathcal{M}_\pm^\pm&=(N^ie_i)^\pm \pm \frac{1}{2}\left(r^i(\nabla_iN^j)e_j\right)^\pm \pm \frac{1}{2} \left(R^l_{\,ikj}r^ir^jN^k\frac{\partial}{\partial r^l}\right)^\pm+\mathcal{O}(r^2)^\pm \nonumber \\
&=\pm N^i\frac{\partial}{\partial r^i}+\frac{1}{2}N^ie_i+\frac{1}{2}r^i(\nabla_i N^j)\frac{\partial}{\partial r^j}+\mathcal{O}(r)
\end{align}
\begin{align}
\mathcal{M}_+^+ + \mathcal{M}_-^-&=N^ie_i + r^i(\nabla_iN^j)\frac{\partial}{\partial r^j} +\mathcal{O}(r)
\end{align}

\begin{align}
\mathcal{T}_\pm&=\mp\frac{1}{2}[r_E, \partial_t]-\frac{1}{8}[r_E, [r_E, \partial_t]]+\mathcal{O}(r^3) \nonumber \\
&=\mp\frac{1}{2}\dot{\Gamma}^k_{ij}r^ir^j\frac{\partial}{\partial r^k}+\frac{1}{8}\left(\dot{\Gamma}^k_{ij}r^ir^je_k-r^i(\tilde{\nabla}_i\dot{\Gamma}^l_{jk})r^jr^k\frac{\partial}{\partial r^l}\right)+\mathcal{O}(r^3)
\end{align}

\begin{align}
\mathcal{T}_\pm^\pm&=\mp\frac{1}{2}\left(\dot{\Gamma}^k_{ij}r^ir^j\frac{\partial}{\partial r^k}\right)^\pm+\frac{1}{8}\left(\dot{\Gamma}^k_{ij}r^ir^je_k-r^i(\tilde{\nabla}_i\dot{\Gamma}^l_{jk})r^jr^k\frac{\partial}{\partial r^l}\right)^\pm+\mathcal{O}(r^3)^\pm \nonumber \\
&=\mp\frac{1}{8}\dot{\Gamma}^k_{ij}r^ir^j\frac{\partial}{\partial r^k}+\mathcal{O}(r^2)
\end{align}

\begin{align}
\mathcal{T}_+^+ + \mathcal{T}_-^-&=\mathcal{O}(r^2)
\end{align}

Note that
\begin{align}
\sqrt{\gamma}\int d^3r\, e^{-\frac{i}{\hbar}r^ip_i} r^j \frac{\partial}{\partial r^k}\left\langle:f^+g^-:\right\rangle&=-\left(p_k\frac{\partial}{\partial p_j}+\delta^j_k\right)\sqrt{\gamma}\int d^3r\, e^{-\frac{i}{\hbar}r^ip_i}\left\langle:f^+g^-:\right\rangle, \\
\sqrt{\gamma}\int d^3r\, e^{-\frac{i}{\hbar}r^ip_i} e_k\left\langle:f^+g^-:\right\rangle&=D_k\sqrt{\gamma}\int d^3r\, e^{-\frac{i}{\hbar}r^ip_i}\left\langle:f^+g^-:\right\rangle,
 \label{eqn:ek_on_f+g-}
\end{align}
where
\begin{equation}
\label{eq:D_i}
D_i:=\nabla_i+\Gamma^k_{ij}p_k\frac{\partial}{\partial p_j},
\end{equation}
and $D_ip_j=0$. Using the above results,
\begin{align} \label{eqn:M+M+T+T_on_f+g-}
\sqrt{\gamma}&\int d^3r\, e^{-\frac{i}{\hbar}r^ip_i} (\mathcal{M}_+^+ + \mathcal{M}_-^- + \mathcal{T}_+^+ + \mathcal{T}_-^-)\left\langle:f^+g^-:\right\rangle \nonumber \\
&=\sqrt{\gamma}\int d^3r\, e^{-\frac{i}{\hbar}r^ip_i} \left[N^ie_i+r^i(\nabla_iN^j)\frac{\partial}{\partial r^j}+\mathcal{O}(r)\right]\left\langle:f^+g^-:\right\rangle \nonumber \\
&=\left[N^iD_i-(\nabla_iN^j)\left(p_j\frac{\partial}{\partial p_i}+\delta^i_j\right)+\mathcal{O}(\hbar)\right]\sqrt{\gamma}\int d^3r\, e^{-\frac{i}{\hbar}r^ip_i} \left\langle:f^+g^-:\right\rangle
\end{align}

Next consider the terms with $\Delta_*^\pm$,
\begin{align}
\label{eq:Delta^p-Delta^m}
(\Delta_*^+-\Delta_*^-)(f^+g^-)&=(\Delta f)^+g^--f^+(\Delta g)^- \nonumber \\
&=(\tilde{\Delta} f^+)g^--f^+(\tilde{\Delta}g^-)-\left([\tilde{\Delta},e^{\frac{r^i}{2}\tilde{\nabla}_i}]f\right)g^-+f^+\left([\tilde{\Delta},e^{-\frac{r^i}{2}\tilde{\nabla}_i}]g\right),
\end{align}
where $\tilde{\Delta}=\gamma^{ij}\tilde{\nabla}_i\tilde{\nabla}_j$. Here we use a slightly different reasoning from Ref.~\cite{Friedrich:2018qjv}. Since \eqref{eq:P^pmf^mp} holds for any $X^i$, we have 
\begin{equation}
\label{eq:convert}
\tilde{\nabla}_if^\pm=\left(\pm 2\frac{\partial}{\partial r^i}+\mathcal{O}(r)\right)f^\pm.
\end{equation}
Then we have
\begin{equation} \label{eqn:formula_z=1}
(\tilde{\Delta} f^+)g^- -f^+(\tilde{\Delta} g^-)=\left( 2\gamma^{ij}\tilde{\nabla}_i \frac{\partial}{\partial r^j}+\mathcal{O}(r)\right)(f^+g^-).
\end{equation}
Noting that $[\tilde{\Delta},e^{\pm\frac{r^i}{2}\tilde{\nabla}_i}]f=\mathcal{O}(r)f^\pm$, the remaining terms in \eqref{eq:Delta^p-Delta^m} are $\mathcal{O}(r^0)\langle:f^+g^-:\rangle$. Therefore
\begin{align} \label{eqn:integratedformula_z=1}
\sqrt{\gamma}\int d^3r\, e^{-\frac{i}{\hbar}r^ip_i}(\Delta_*^+-\Delta_*^-)\left\langle:f^+g^-:\right\rangle&=\left(\frac{2i}{\hbar}\gamma^{ij}p_iD_j +\mathcal{O}(\hbar^0)\right)\sqrt{\gamma}\int d^3r\, e^{-\frac{i}{\hbar}r^ip_i}\left\langle:f^+g^-:\right\rangle
\end{align}
Recalling \eqref{eq:del_tF},
\begin{align}
&\partial_t F_{\phi\phi}=N(F_{\phi_\perp\phi}+F_{\phi\phi_\perp})+\left(N^iD_i-(\nabla_iN^j)p_j\frac{\partial}{\partial p_i}+NK\right)F_{\phi\phi}+\mathcal{O}(\hbar)(F_{\phi_\perp\phi}-F_{\phi\phi_\perp}) \label{eqn:dFphiphidt} \\
&\partial_t F_{\phi_\perp\phi_\perp}=\left(N^iD_i-(\nabla_iN^j)p_j\frac{\partial}{\partial p_i}-NK+\mathcal{O}(\hbar)\right)F_{\phi_\perp\phi_\perp}-\left(\frac{i}{\hbar}Np_iD^i+\mathcal{O}(\hbar^0)\right)(F_{\phi_\perp\phi}-F_{\phi\phi_\perp})+\mathcal{O}(F_{\phi_\perp\phi}+F_{\phi\phi_\perp}) \\
&\frac{i}{2}\partial_t (F_{\phi_\perp\phi}-F_{\phi\phi_\perp})=\frac{i}{2}\left(N^iD_i-(\nabla_iN^j)p_j\frac{\partial}{\partial p_i}+\mathcal{O}(\hbar)\right)(F_{\phi_\perp\phi}-F_{\phi\phi_\perp})-\left(\frac{1}{\hbar}Np_iD^i+\mathcal{O}(\hbar^0)\right)F_{\phi\phi}+\mathcal{O}(\hbar)(F_{\phi_\perp\phi}+F_{\phi\phi_\perp})
\end{align}
Taking linear combinations \eqref{eq:f_1^p}--\eqref{eq:f_3},
\begin{align}
\partial_t f_1^+&=\left(N^iD_i-(\nabla_iN^j)p_j\frac{\partial}{\partial p_i}\right)f_1^+-N\frac{p_i}{\omega}D^if_1^-+\mathcal{O}(f_2,f_3,\hbar) \\
\partial_t f_1^-&=\left(N^iD_i-(\nabla_iN^j)p_j\frac{\partial}{\partial p_i}\right)f_1^--N\frac{p_i}{\omega}D^if_1^++\mathcal{O}(f_2,f_3,\hbar)
\end{align}
Therefore, the equation for $f_1=f_1^++f_1^-$ is
\begin{equation} 
\left[\partial_t - N^iD_i + (\nabla_iN^j)p_j\frac{\partial}{\partial p_i} +N\frac{p_i}{\omega}D^i  \right]f_1=\mathcal{O}(f_2,f_3,\hbar)\,. \label{eqn:kineticeq_relativistic}
\end{equation}
This coincides with the Boltzmann equation by undoing the $3+1$ decomposition,
\begin{equation}
\left[ p^{(4)\mu} \partial_\mu +\Gamma^{(4)\nu}_{\mu i}p^{(4)\mu} p^{(4)}_\nu \frac{\partial}{\partial p_i}\right]f_1\simeq 0
\end{equation}
where $p^{(4)}_i=p_i$, $p^{(4)}_0=-N\omega +N^ip_i$ so that $g^{\mu\nu}p^{(4)}_\mu p^{(4)}_\nu=0$, $g_{\mu\nu}$ is the spacetime metric, and $\Gamma^{(4)\nu}_{\mu \rho}$ is the Christoffel symbols composed of $g_{\mu\nu}$. We refer readers to Ref.~\cite{Friedrich:2018qjv} for higher order terms in $\hbar$, self-interaction terms of the scalar field, and spatial dependence of the lapse function.

\section{Kinetic equation for Lifshitz scalar}
\label{sec:Lifshitz scalar}

In this section we study the kinetic equation for a real scalar field with a general dispersion relation, whose action is
\begin{equation}
\label{eq:LifshitzAction}
 I=\frac{1}{2}\int dtd^3x\, N\sqrt{\gamma}\left[ \left(\partial_\perp\phi \right)^2 + \phi \mathcal{F}(\Delta)\phi \right]\,,
\end{equation}
where $\mathcal{F}(\Delta)$ is a function specified by the theory. We still impose the projectability condition, i.e., $N=N(t)$. (As already mentioned in Sec.~\ref{sec:intro}, the renormalizability of HL gravity was proven under the projectability condition~\cite{Barvinsky:2015kil,Barvinsky:2017zlx}.) The equation of motion is
\begin{equation}
\label{eq:LifshitzEOM}
 \partial_\perp^2\phi+K\partial_\perp\phi - \mathcal{F}(\Delta)\phi = 0\,.
\end{equation}

Again, we are interested in the evolution equation of $f_1 = f_1^+ + f_1^-$ for the same reason as discussed just after Eqs.~\eqref{eq:E}--\eqref{eq:Sij} (see Appendixes \ref{sec:smallness} and \ref{sec:energy-momentm}), where $f_1^{\pm}$ are given by (\ref{eq:f_1^p}) and (\ref{eq:f_1^m}) with
\begin{equation}
 \omega = \hbar \sqrt{-\mathcal{F}(-p^2/\hbar^2)}\,, \quad p^2 = \gamma^{ij}p_ip_j\,.
\end{equation}
We can still use (\ref{eq:del_tF}) to evaluate time derivatives of the Wigner functions, and thus we simply need to compute $\sqrt{\gamma}\partial_t(F_{\phi\phi}/\sqrt{\gamma})$, $\sqrt{\gamma}\partial_t((F_{\phi_\perp\phi}-F_{\phi\phi_\perp})/\sqrt{\gamma})$ and $\sqrt{\gamma}\partial_t((F_{\phi_\perp\phi_\perp}/\sqrt{\gamma})$. The computation of $\sqrt{\gamma}\partial_t(F_{\phi\phi}/\sqrt{\gamma})$ in the previous section does not rely on the equation of motion and thus is independent of the dispersion relation. We are thus able to use the same formula (\ref{eqn:dFphiphi}). On the other hand, the computations of the other two quantities rely on the equation of motion, and thus we need to reconsider them. Using (\ref{eq:LifshitzEOM}) instead of (\ref{eqn:eom_relativistic}) but still assuming $N=N(t)$, we obtain
\begin{align}
\frac{1}{2}\sqrt{\gamma}\partial_t((F_{\phi_\perp\phi}-F_{\phi\phi_\perp})/\sqrt{\gamma})=&\frac{1}{2}\sqrt{\gamma}\int d^3r\, e^{-\frac{i}{\hbar}r^ip_i} (\mathcal{T}_+^+ + \mathcal{T}_-^- + \mathcal{M}_+^+ + \mathcal{M}_-^--\frac{1}{2}(NK)^+-\frac{1}{2}(NK)^-)\left\langle:\phi_\perp^+\phi^- - \phi^+\phi_\perp^-:\right\rangle \nonumber \\
&-\frac{1}{2}\sqrt{\gamma}\int d^3r\, e^{-\frac{i}{\hbar}r^ip_i} (\frac{1}{2}(NK)^+-\frac{1}{2}(NK)^-)\left\langle:\phi_\perp^+\phi^- + \phi^+\phi_\perp^-:\right\rangle \nonumber \\
 &+\frac{1}{2}\sqrt{\gamma}\int d^3r\,e^{-\frac{i}{\hbar}r^ip_i}((N\mathcal{F}(\Delta))_*^+-(N\mathcal{F}(\Delta))_*^-)\left\langle:\phi^+\phi^-:\right\rangle\,,
 \label{eqn:dFphiperpphi-dFphiphiperp_general}
\end{align}
and 
\begin{align}
\sqrt{\gamma}\partial_t((F_{\phi_\perp\phi_\perp}/\sqrt{\gamma})=&\sqrt{\gamma}\int d^3r\, e^{-\frac{i}{\hbar}r^ip_i} (\mathcal{T}_+^+ + \mathcal{T}_-^- + \mathcal{M}_+^+ + \mathcal{M}_-^--(NK)^+-(NK)^-)\left\langle:\phi_\perp^+\phi_\perp^-:\right\rangle \nonumber \\
&-\frac{1}{2}\sqrt{\gamma}\int d^3r\,e^{-\frac{i}{\hbar}r^ip_i}((N\mathcal{F}(\Delta))_*^+-(N\mathcal{F}(\Delta))_*^-)\left\langle:\phi_\perp^+\phi^- - \phi^+\phi_\perp^-:\right\rangle \nonumber \\
&+\frac{1}{2}\sqrt{\gamma}\int d^3r\,e^{-\frac{i}{\hbar}r^ip_i}((N\mathcal{F}(\Delta))_*^++(N\mathcal{F}(\Delta))_*^- )\left\langle:\phi_\perp^+\phi^- + \phi^+\phi_\perp^-:\right\rangle\,.
 \label{eqn:dFphiperpphiperp_general}
\end{align}
The only difference between these expressions and those in the previous section is that $\Delta$ is now replaced by $\mathcal{F}(\Delta)$. The remaining computation proceeds almost identically as in the previous section, except that we now need to deal with
\begin{equation}
\label{eq:F++F-}
 (\mathcal{F}(\Delta)^+_*-\mathcal{F}(\Delta)^-_*)(f^+g^-)
  = (\mathcal{F}(\tilde{\Delta})f^+)g^- -  f^+(\mathcal{F}(\tilde{\Delta})g^-) + \mathcal{O}(r) f^+g^-\,.
\end{equation}
We thus suppose that the leading order expression on the right-hand side of this equation is expanded as
\begin{equation} \label{eqn:ansatz}
 (\mathcal{F}(\tilde{\Delta})f^+)g^- -  f^+(\mathcal{F}(\tilde{\Delta})g^-)
   = \sum_{n,m}\sum_{\{i\}, \{j\}} c^{i_1\cdots i_nj_1\cdots j_m}\tilde{\nabla}_{i_1}\cdots\tilde{\nabla}_{i_n}\frac{\partial}{\partial r^{j_1}}\cdots \frac{\partial}{\partial r^{j_m}}(f^+g^-)\,,
\end{equation}
where $\{c^{i_1\cdots i_nj_1\cdots j_m}\}$ is a set of coefficients that are independent of $r^k$. If the function $\mathcal{F}(\Delta)$ is given as a linear combination of several terms, then one can simply compute $\{c^{i_1\cdots i_nj_1\cdots j_m}\}$ for each term and then form the linear combination. It is obvious that a constant in $\mathcal{F}(\Delta)$ gives zero contribution to $\{c^{i_1\cdots i_nj_1\cdots j_m}\}$. For $\Delta\in \mathcal{F}(\Delta)$, we already have the formula (\ref{eqn:formula_z=1}). In the following subsections, we shall consider $\Delta^2\in\mathcal{F}(\Delta)$ and $\Delta^3\in\mathcal{F}(\Delta)$. (In the latter case, for technical simplicity we shall restrict our consideration to spatially flat backgrounds.)

Once the set of coefficients $\{c^{i_1\cdots i_nj_1\cdots j_m}\}$ is computed for a given function $\mathcal{F}(\Delta)$, it is easy to obtain
\begin{align}
 \sqrt{\gamma}\int d^3r\, e^{-\frac{i}{\hbar}r^ip_i}(\mathcal{F}(\Delta)_*^+-\mathcal{F}(\Delta)_*^-)\left\langle:f^+g^-:\right\rangle = &
 (1+\mathcal{O}(\hbar)) \frac{2i}{\hbar}\mathcal{D} \sqrt{\gamma}\int d^3r\, e^{-\frac{i}{\hbar}r^ip_i}\left\langle:f^+g^-:\right\rangle
 \label{eqn:integratedformula_general}
\end{align}
by integration by part and the use of (\ref{eqn:ek_on_f+g-}), where
\begin{equation} \label{eqn:def-calD}
 \mathcal{D}  =  \frac{1}{2}\sum_{n,m}\left(\frac{i}{\hbar}\right)^{m-1}\sum_{\{i\}, \{j\}} c^{i_1\cdots i_nj_1\cdots j_m} D_{i_1}\cdots D_{i_n} p_{j_1}\cdots p_{j_m}\,,
\end{equation}
and $D_i$ is defined in Eq.~\eqref{eq:D_i}. The formula (\ref{eqn:integratedformula_general}) is an obvious generalization of (\ref{eqn:integratedformula_z=1}). As shown in the previous section, by using (\ref{eq:del_tF}) and (\ref{eqn:M+M+T+T_on_f+g-}), (\ref{eqn:dFphiphi}) is rewritten as (\ref{eqn:dFphiphidt}). By using (\ref{eq:del_tF}), (\ref{eqn:M+M+T+T_on_f+g-}) and the new formula (\ref{eqn:integratedformula_general}), Eq.~(\ref{eqn:dFphiperpphi-dFphiphiperp_general}) is written as 
\begin{align} \label{eqn:dFphiperpphi-dFphiphiperpdt_general}
&\frac{i}{2}\partial_t (F_{\phi_\perp\phi}-F_{\phi\phi_\perp})=\frac{i}{2}\left(N^iD_i-(\nabla_iN^j)p_j\frac{\partial}{\partial p_i}+\mathcal{O}(\hbar)\right)(F_{\phi_\perp\phi}-F_{\phi\phi_\perp})-\frac{N}{\hbar}(1+\mathcal{O}(\hbar)) \mathcal{D}F_{\phi\phi}+\mathcal{O}(F_{\phi_\perp\phi}+F_{\phi\phi_\perp})\,.
\end{align}
By using the same set of formulas, Eq.~(\ref{eqn:dFphiperpphiperp_general}) is rewritten as
\begin{align} \label{eqn:dFphiperpphiperpdt_general}
 &\partial_t F_{\phi_\perp\phi_\perp}=\left(N^iD_i-(\nabla_iN^j)p_j\frac{\partial}{\partial p_i}-NK+\mathcal{O}(\hbar)\right)F_{\phi_\perp\phi_\perp} - \frac{iN}{\hbar} (1+\mathcal{O}(\hbar)) \mathcal{D}(F_{\phi_\perp\phi}-F_{\phi\phi_\perp})+\mathcal{O}(F_{\phi_\perp\phi}+F_{\phi\phi_\perp})\,.
\end{align}
It is now straightforward to take linear combinations of (\ref{eqn:dFphiphidt}), (\ref{eqn:dFphiperpphi-dFphiphiperpdt_general}), and (\ref{eqn:dFphiperpphiperpdt_general}) to obtain
\begin{align}
 \partial_t f_1^+&=\left(N^iD_i-(\nabla_iN^j)p_j\frac{\partial}{\partial p_i}\right)f_1^+ - \frac{N}{\omega}\mathcal{D} f_1^- +\mathcal{O}(f_2,f_3,\hbar)\,, \\
 \partial_t f_1^-&=\left(N^iD_i-(\nabla_iN^j)p_j\frac{\partial}{\partial p_i}\right)f_1^- - \frac{N}{\omega}\mathcal{D} f_1^+ +\mathcal{O}(f_2,f_3,\hbar)\,.
\end{align}
Therefore, the equation for $f_1=f_1^++f_1^-$ is
\begin{equation} \label{eqn:kineticeq_general}
 \left[\partial_t - N^iD_i + (\nabla_iN^j)p_j\frac{\partial}{\partial p_i}
+ \frac{N}{\omega}\mathcal{D} \right]f_1=\mathcal{O}(f_2,f_3,\hbar) \simeq 0\,.
\end{equation}

\subsection{$z=2$}

Let us consider the terms up to $z=2$, namely,
\begin{equation} \label{eq:z=2dispersion}
 \mathcal{F}(\Delta) = -\beta \frac{\hbar^2}{M^2}\Delta^2 + c_{\phi}^2 \Delta - \frac{m^2}{\hbar^2}\,,
\end{equation}
where $M$ is the Lifshitz scale, $\beta$ is a dimensionless constant, $c_{\phi}$ is the speed limit at low energy below $M$, and $m$ is the mass. We already know how to deal with the last two terms. Let us therefore focus on the contribution from $\Delta^2$.

We need to reexpress $(\tilde{\Delta}^2 f^+) g^- - f^+ (\tilde{\Delta}^2 g^-)$ in the form of the right-hand side of (\ref{eqn:ansatz}). For this purpose, we first compute several relevant terms on the right-hand side of (\ref{eqn:ansatz}) and then invert the obtained relations. Using \eqref{eq:convert}, at the leading order of $r^i$ it is straightforward to show that 
\begin{equation}
\begin{pmatrix}
24 \gamma^{ij}\gamma^{kl}\tilde{\nabla}_i \frac{\partial}{\partial r^j} \frac{\partial}{\partial r^k} \frac{\partial}{\partial r^l} (f^+g^-) \\
2 \gamma^{ij}\gamma^{kl}\tilde{\nabla}_i \tilde{\nabla}_j \tilde{\nabla}_k \frac{\partial}{\partial r^l} (f^+g^-) \\
2 \gamma^{ik}\gamma^{jl}\tilde{\nabla}_i \tilde{\nabla}_j \tilde{\nabla}_k \frac{\partial}{\partial r^l} (f^+g^-) \\
2 \gamma^{il}\gamma^{jk}\tilde{\nabla}_i \tilde{\nabla}_j \tilde{\nabla}_k \frac{\partial}{\partial r^l} (f^+g^-)
\end{pmatrix}
\simeq
\begin{pmatrix}
2  & 2 & -4 & 1 \\
-2 & 0 & 0  & 1 \\
-1 & 1 & 1  & 0 \\
1  & 1 & 3  & 0
\end{pmatrix}
\begin{pmatrix}
(\tilde{\nabla}^if^+)(\tilde{\nabla}_i\tilde{\Delta} g^-) - (\tilde{\nabla}_i\tilde{\Delta} f^+)(\tilde{\nabla}^ig^-) \\
(\tilde{\nabla}_i\tilde{\nabla}_j\tilde{\nabla}^i\tilde{\nabla}^jf^+)g^- - f^+(\tilde{\nabla}_i\tilde{\nabla}_j\tilde{\nabla}^i\tilde{\nabla}^jg^-) \\
(\tilde{\Delta}\tilde{\nabla}_if^+)(\tilde{\nabla}^i g^-) - 
(\tilde{\nabla}^i f^+)(\tilde{\Delta}\tilde{\nabla}_ig^-) \\
(\tilde{\Delta}^2 f^+)g^- - f^+(\tilde{\Delta}^2 g^-)
\end{pmatrix}\,.
\end{equation}
Inverting the coefficient matrix, we then obtain 
\begin{equation}
(\tilde{\Delta}^2 f^+)g^- - f^+(\tilde{\Delta}^2 g^-) \simeq \left[ 4\gamma^{i(j}\gamma^{kl)}\tilde{\nabla}_i \frac{\partial}{\partial r^j} \frac{\partial}{\partial r^k} \frac{\partial}{\partial r^l}+\left( \frac{5}{3}\gamma^{ij}\gamma^{kl} - \frac{5}{3}\gamma^{ik}\gamma^{jl} + \gamma^{il}\gamma^{jk}  \right) \tilde{\nabla}_i \tilde{\nabla}_j \tilde{\nabla}_k \frac{\partial}{\partial r^l} \right](f^+g^-)\,.
\end{equation}
Since the first term in the right-hand side is the leading order in $\hbar$ expansion, we have
\begin{equation}
 (\mathcal{F}(\tilde{\Delta}) f^+)g^- - f^+(\mathcal{F}(\tilde{\Delta}) g^-) \simeq \left[ 2c_{\phi}^2 \gamma^{ij}\tilde{\nabla}_i\frac{\partial}{\partial r^j} - 4\beta \frac{\hbar^2}{M^2}\gamma^{ij}\gamma^{kl}\tilde{\nabla}_i \frac{\partial}{\partial r^j} \frac{\partial}{\partial r^k} \frac{\partial}{\partial r^l}\right](f^+g^-)\,.
\end{equation}
Therefore the operator $\mathcal{D}$ defined in (\ref{eqn:def-calD}) is
\begin{equation} \label{eqn:calD_z=2}
 \mathcal{D} = c_{\phi}^2 p_iD^i + 2\beta \frac{p^2}{M^2}p_iD^i\,,
\end{equation}
and the kinetic equation is as shown in (\ref{eqn:kineticeq_general}), where
\begin{equation} \label{eqn:omega_z=2}
 \omega = \sqrt{\frac{\beta}{M^2}p^4 + c_{\phi}^2 p^2 + m^2}\,, \quad
  p^4 = (p^2)^2\,, \quad p^2 = \gamma^{ij}p_ip_j\,,
\end{equation}
and $D_i$ is defined in Eq.~\eqref{eq:D_i}. This result is valid in general curved geometries with the projectability condition, $N=N(t)$.

\subsection{$z=3$, spatially flat}
\label{sec:z=3flat}

Next we consider terms up to $z=3$, namely,
\begin{equation} \label{eqn:F(Delta)_z=3}
 \mathcal{F}(\Delta) = \alpha \frac{\hbar^4}{M^4}\Delta^3 -\beta \frac{\hbar^2}{M^2}\Delta^2 + c_{\phi}^2 \Delta - \frac{m^2}{\hbar^2}\,,
\end{equation}
where $\alpha$ is a dimensionless constant. Since we already know how to deal with the last three terms, let us focus on the contribution from $\Delta^3$. We need to reexpress $(\tilde{\Delta}^3 f^+) g^- - f^+ (\tilde{\Delta}^3 g^-)$ in the form of the right-hand side of (\ref{eqn:ansatz}). For this purpose, again, we first compute several relevant terms on the right-hand side of (\ref{eqn:ansatz}) and then invert the obtained relations. In this case, however, there are much more relevant terms than in the previous case. For simplicity let us therefore restrict our consideration to spatially flat geometries, i.e. $\gamma_{ij}=a(t)^2\delta_{ij}$, where $a$ is an arbitrary positive function of the time variable $t$. This significantly simplifies the computation but still allows for several interesting examples including the flat Friedmann-Lema\^{i}tre-Robertson-Walker geometry and spherically symmetric geometries in the (generalized) Painlev\'{e}-Gullstrand coordinate system~\cite{Painleve,Gullstrand:1922tfa,DeFelice:2018vza}.

Using \eqref{eq:convert}, at the leading order of $r^i$ it is straightforward to show that 
\begin{equation}
\begin{pmatrix}
32 \gamma^{ij}\gamma^{kl}\gamma^{mn}\partial_i \frac{\partial}{\partial r^j} \frac{\partial}{\partial r^k} \frac{\partial}{\partial r^l} \frac{\partial}{\partial r^m} \frac{\partial}{\partial r^n} (f^+g^-) \\
8 \gamma^{ij}\gamma^{kl}\gamma^{mn}\partial_i \partial_j \partial_k \frac{\partial}{\partial r^l} \frac{\partial}{\partial r^m} \frac{\partial}{\partial r^n} (f^+g^-) \\
8 \gamma^{il}\gamma^{jm}\gamma^{nk}\partial_i \partial_j \partial_k \frac{\partial}{\partial r^l} \frac{\partial}{\partial r^m} \frac{\partial}{\partial r^n} (f^+g^-) \\
2 \gamma^{ij}\gamma^{kl}\gamma^{mn}\partial_i \partial_j \partial_k \partial_l \partial_m \frac{\partial}{\partial r^n} (f^+g^-)
\end{pmatrix}
=
\begin{pmatrix}
-4 & 4 & 1 & 1 \\
0 & -4 & 1 & 1 \\
0 & 0 & -3 & 1 \\
4 & 4 & 1 & 1
\end{pmatrix}
\begin{pmatrix}
(\Delta^2\partial_if^+)(\partial^i g^-) - (\partial^i f^+)(\Delta^2\partial_i g^-)\\
(\Delta\partial^i\partial^j f^+)(\partial_i\partial_j g^-) - (\partial_i\partial_j f^+)(\Delta\partial^i\partial^j g^-) \\
(\Delta^2 f^+)(\Delta g^-) - 
(\Delta f^+)(\Delta^2 g^-) \\
(\Delta^3 f^+)g^- - f^+(\Delta^3 g^-)
\end{pmatrix}\,.
\end{equation}
Inverting the coefficient matrix, we then obtain
\begin{align}
(\tilde{\Delta}^3 f^+)g^- - f^+(\tilde{\Delta}^3 g^-) = &\biggl[ 6\gamma^{ij}\gamma^{kl}\gamma^{mn}\partial_i \frac{\partial}{\partial r^j} \frac{\partial}{\partial r^k} \frac{\partial}{\partial r^l} \frac{\partial}{\partial r^m} \frac{\partial}{\partial r^n} \nonumber \\
& + \left( 3\gamma^{ij}\gamma^{kl}\gamma^{mn} + 2\gamma^{il}\gamma^{jm}\gamma^{nk} \right) \partial_i \partial_j \partial_k \frac{\partial}{\partial r^l} \frac{\partial}{\partial r^m} \frac{\partial}{\partial r^n} \nonumber \\
&+ \frac{3}{8}\gamma^{ij}\gamma^{kl}\gamma^{mn}\partial_i \partial_j \partial_k \partial_l \partial_m \frac{\partial}{\partial r^n} \biggr](f^+g^-)
\end{align}
Since the first term in the right-hand side is the leading order in $\hbar$ expansion, we have
\begin{eqnarray}
 (\mathcal{F}(\tilde{\Delta}) f^+)g^- - f^+(\mathcal{F}(\tilde{\Delta}) g^-) & \simeq & \left[ 2c_{\phi}^2 \gamma^{ij}\tilde{\nabla}_i\frac{\partial}{\partial r^j} - 4\beta \frac{\hbar^2}{M^2}\gamma^{ij}\gamma^{kl}\tilde{\nabla}_i \frac{\partial}{\partial r^j} \frac{\partial}{\partial r^k} \frac{\partial}{\partial r^l} \right. \nonumber \\
 & & \quad \left. + 6\alpha \frac{\hbar^4}{M^4} \gamma^{ij}\gamma^{kl}\gamma^{mn}\tilde{\nabla}_i \frac{\partial}{\partial r^j} \frac{\partial}{\partial r^k} \frac{\partial}{\partial r^l} \frac{\partial}{\partial r^m} \frac{\partial}{\partial r^n} \right](f^+g^-)\,.
\end{eqnarray}
Therefore the operator $\mathcal{D}$ defined in (\ref{eqn:def-calD}) is
\begin{equation} \label{eqn:calD_z=3}
 \mathcal{D} = c_{\phi}^2 p_iD^i + 2\beta \frac{p^2}{M^2}p_iD^i + 3\alpha \frac{p^4}{M^2}p_iD^i\,,
\end{equation}
and the kinetic equation is as shown in (\ref{eqn:kineticeq_general}), where
\begin{equation} \label{eqn:omega_z=3}
 \omega = \sqrt{\frac{\alpha}{M^4}p^6 + \frac{\beta}{M^2}p^4 + c_{\phi}^2 p^2 + m^2}\,, \quad p^6 = (p^2)^3\,, \quad p^4 = (p^2)^2\,, \quad p^2 = \gamma^{ij}p_ip_j\,,
\end{equation}
and $D_i$ is defined in Eq.~\eqref{eq:D_i}. This result is valid in spatially flat geometries, $\gamma_{ij} = a(t)^2\delta_{ij}$, with the projectability condition, $N=N(t)$.

\subsection{General $\mathcal{F}(\Delta)$}

Based on the results of the previous subsections, we conjecture that the kinetic equation for general $\mathcal{F}(\Delta)$ in general curved geometries with the projectability condition, $N=N(t)$, should be given by (\ref{eqn:kineticeq_general}) with 
\begin{equation}
 \mathcal{D} = \mathcal{F}'(-p^2/\hbar^2)p_iD^i\,,  \quad
   \omega = \hbar \sqrt{-\mathcal{F}(-p^2/\hbar^2)}\,,
\end{equation}
where $p^2 = \gamma^{ij}p_ip_j$. The conjectured kinetic equation for the general dispersion relation can be rewritten as 
\begin{equation} \label{eqn:kineticeq_general_vg}
 \left[\partial_t - N^iD_i + (\nabla_iN^j)p_j\frac{\partial}{\partial p_i}
  + N v_{\rm g}^iD_i \right]f_1=\mathcal{O}(f_2,f_3,\hbar) \simeq 0\,,
\end{equation}
where
\begin{equation} \label{eq:groupvelocity}
 v_{\rm g}^i \equiv \frac{\partial\omega}{\partial p_i}
\end{equation}
is the group velocity.

\section{Summary and Discussion}
\label{sec:summary}

In the present paper we have derived classical kinetic (Boltzmann-like) equations of distribution functions for a real scalar field with the Lifshitz scaling $z=2$ on curved space and $z=3$ on spatially flat space under the assumption of the projectability condition, $N=N(t)$. We have then conjectured that (\ref{eqn:kineticeq_general_vg}) gives the kinetic equation for a real scalar field with a general dispersion relation in general curved geometries satisfying the projectability condition. In the kinetic equation all the information about the nontrivial dispersion relation is included in the expression of the group velocity $v_{\rm g}^i$. This in particular implies that the free streaming of collisionless particles and the diffusion damping should occur with the group velocity. This agrees with a natural expectation from the well-known fact that the peak of a wave packet propagates with the group velocity.

We have employed the method of Wigner functions on curved space based on Ref.~\cite{Friedrich:2018qjv}. The application of the method to a Lifshitz scalar was made possible by noting the conversion between derivatives $\tilde{\nabla}_i$ and $\partial/\partial r^i$ \eqref{eq:convert} at the leading order of expansion in $\hbar$. What we have not done in the present paper is to provide a proof of the conjectured form of the kinetic equation (\ref{eqn:kineticeq_general_vg}), while it correctly reproduces (\ref{eqn:kineticeq_relativistic}) for the relativistic case, (\ref{eqn:kineticeq_general}) with (\ref{eqn:calD_z=2}) and (\ref{eqn:omega_z=2}) for $z=2$, and (\ref{eqn:kineticeq_general}) with (\ref{eqn:calD_z=3}) and (\ref{eqn:omega_z=3}) for $z=3$. It is also interesting to obtain terms higher order in $\hbar$, spatial dependence of the lapse function, and collision terms.

As already mentioned in the Introduction, the Lifshitz scaling is at the heart of the renormalizability of HL gravity. Moreover, the same scaling has significant implications to cosmology. For example, the Lifshitz scaling with $z=3$ can solve the horizon problem and lead to a novel mechanism for generation of almost scale-invariant cosmological perturbations~\cite{Mukohyama:2009gg}. Moreover, the Lifshitz scaling can solve the flatness problem without inflation~\cite{Bramberger:2017tid}. Armed with the method developed in the present paper, we now aim to tackle the isotropy problem \cite{Albrecht:1998ir} in the HL gravity, which will be addressed in the near future.

\section*{Acknowledgments}
This work was supported in part by Japan Society for the Promotion of Science (JSPS) Grants-in-Aid for Scientific Research (KAKENHI) No.\ JP17H02890 (S.\,M.), No.\ JP17H06359 (S.\,M.), and No.\ 16J06266 (Y.\,W.) as well as by World Premier International Research Center Initiative (WPI), MEXT, Japan. Y.\,W.'s work was supported in part by the Program for Leading Graduate Schools, Ministry of Education, Culture, Sports, Science and Technology (MEXT), Japan. Y.\,W.\ is grateful to Yukawa Institute for Theoretical Physics at Kyoto University for warm hospitality.

\appendix

\section{Smallness of other Wigner functions $f_2$ and $f_3$}
\label{sec:smallness}

This appendix section is devoted to a discussion that $f_2$ and $f_3$ remain small if they are set so initially. Although we essentially follow the arguments in Ref.~\cite{Friedrich:2018qjv} for $z=1$, we assume the general dispersion relation Eq.~\eqref{eq:LifshitzAction} with the projectability condition $N=N(t)$. To derive time derivatives of $f_2$, $f_3$, we evaluate
\begin{align}
\frac{1}{2}\sqrt{\gamma}\partial_t((F_{\phi_\perp\phi}+F_{\phi\phi_\perp})/\sqrt{\gamma})=&N\sqrt{\gamma}\int d^3r\, e^{-\frac{i}{\hbar}r^ip_i}\left\langle:\phi_\perp^+\phi_\perp^-:\right\rangle \nonumber \\
&+ \frac{1}{2}\sqrt{\gamma}\int d^3r\, e^{-\frac{i}{\hbar}r^ip_i} (\mathcal{T}_+^+ + \mathcal{T}_-^- + \mathcal{M}_+^+ + \mathcal{M}_-^--\frac{1}{2}(NK)^+-\frac{1}{2}(NK)^-)\left\langle:\phi_\perp^+\phi^- + \phi^+\phi_\perp^-:\right\rangle \nonumber \\
&-\frac{1}{2}\sqrt{\gamma}\int d^3r\, e^{-\frac{i}{\hbar}r^ip_i} (\frac{1}{2}(NK)^+-\frac{1}{2}(NK)^-)\left\langle:\phi_\perp^+\phi^- - \phi^+\phi_\perp^-:\right\rangle \nonumber \\
&+\frac{1}{2}\sqrt{\gamma}\int d^3r\,e^{-\frac{i}{\hbar}r^ip_i}((N\mathcal{F}(\Delta))_*^++(N\mathcal{F}(\Delta))_*^-)\left\langle:\phi^+\phi^-:\right\rangle\,,
\label{eqn:dFphiperpphi+dFphiphiperp_general}
\end{align}
Similar to Eqs.~\eqref{eq:F++F-}, we need to deal with
\begin{equation}
(\mathcal{F}(\Delta)^+_*+\mathcal{F}(\Delta)^-_*)(f^+g^-)
= (\mathcal{F}(\tilde{\Delta})f^+)g^- +  f^+(\mathcal{F}(\tilde{\Delta})g^-) + \mathcal{O}(r) f^+g^-\,.
\end{equation}
As Eq.~\eqref{eqn:ansatz}, we suppose an ansatz
\begin{equation} \label{eq:ansatzSymmetric}
(\mathcal{F}(\tilde{\Delta})f^+)g^- +  f^+(\mathcal{F}(\tilde{\Delta})g^-)
= \sum_{n,m}\sum_{\{i\}, \{j\}} \tilde{c}^{i_1\cdots i_nj_1\cdots j_m}\tilde{\nabla}_{i_1}\cdots\tilde{\nabla}_{i_n}\frac{\partial}{\partial r^{j_1}}\cdots \frac{\partial}{\partial r^{j_m}}(f^+g^-)\,,
\end{equation}
Once the set of coefficients $\{\tilde{c}^{i_1\cdots i_nj_1\cdots j_m}\}$ is computed for a given function $\mathcal{F}(\Delta)$, we can obtain
\begin{align} \label{eqn:integratedformula_general_symmetric}
\sqrt{\gamma}\int d^3r\, e^{-\frac{i}{\hbar}r^ip_i}(\mathcal{F}(\Delta)_*^++\mathcal{F}(\Delta)_*^-)\left\langle:f^+g^-:\right\rangle = &
-\frac{2}{\hbar^2}(1+\mathcal{O}(\hbar))\tilde{\mathcal{D}} \sqrt{\gamma}\int d^3r\, e^{-\frac{i}{\hbar}r^ip_i}\left\langle:f^+g^-:\right\rangle
\end{align}
as Eq.~\eqref{eqn:integratedformula_general}, where
\begin{equation} \label{eqn:def-calDtilde}
\tilde{\mathcal{D}}  =  \frac{1}{2}\sum_{n,m}\left(\frac{i}{\hbar}\right)^{m-2}\sum_{\{i\}, \{j\}} \tilde{c}^{i_1\cdots i_nj_1\cdots j_m} D_{i_1}\cdots D_{i_n} p_{j_1}\cdots p_{j_m}.
\end{equation}
Implementing Eq.~\eqref{eqn:integratedformula_general_symmetric}, the formula Eq.~\eqref{eqn:dFphiperpphi+dFphiphiperp_general} is rewritten as
\begin{align} \label{eqn:dFphiperpphi+dFphiphiperpdt_general}
\frac{1}{2}\partial_t (F_{\phi_\perp\phi}+F_{\phi\phi_\perp})=&N F_{\phi_\perp\phi_\perp} - \frac{N}{\hbar^2}(1+\mathcal{O}(\hbar))\tilde{\mathcal{D}}F_{\phi\phi} + \mathcal{O}(\hbar^0)(F_{\phi_\perp\phi}+F_{\phi\phi_\perp})+\mathcal{O}(\hbar)(F_{\phi_\perp\phi}-F_{\phi\phi_\perp})\,.
\end{align}
This translates into
\begin{equation}
\label{eq:df3dt}
\partial_t f_3=-2\frac{N\omega}{\hbar}f_2 + \mathcal{O}(\hbar^0)(f_1,f_2,f_3)
\end{equation}
if $\tilde{\mathcal{D}}=\omega^2$, which is true for examples considered in the following subsections.

On the other hand, Eq.~\eqref{eqn:dFphiperpphiperpdt_general} is elaborated by
\begin{align}
\partial_t F_{\phi_\perp\phi_\perp}=&\left(N^iD_i-(\nabla_iN^j)p_j\frac{\partial}{\partial p_i}-NK+\mathcal{O}(\hbar)\right)F_{\phi_\perp\phi_\perp} \nonumber \\
&- \frac{iN}{\hbar} (1+\mathcal{O}(\hbar)) \mathcal{D}(F_{\phi_\perp\phi}-F_{\phi\phi_\perp})-\frac{N}{\hbar^2}(1+\mathcal{O}(\hbar))\tilde{\mathcal{D}}(F_{\phi_\perp\phi}+F_{\phi\phi_\perp})\,.
\end{align}
Combining Eq.~\eqref{eqn:dFphiphidt}, we obtain
\begin{equation}
\label{eq:df2dt}
\partial_t f_2=2\frac{N\omega}{\hbar}f_3 + \mathcal{O}(\hbar^0)(f_1,f_2,f_3)
\end{equation}
again if $\tilde{\mathcal{D}}=\omega^2$, which is true for examples considered in the following subsections.

The set of Eqs.~\eqref{eq:df3dt} and \eqref{eq:df2dt} tells us that $f_2$ and $f_3$ are oscillators with the frequency $\sim 2\omega/\hbar$ with respect to the proper time $\int Ndt$ at the leading order in $\hbar$. Therefore, $f_2$ and $f_3$ remain small if they are set to small initially. In the following subsections, let us consider examples where one can explicitly show the relation $\tilde{\mathcal{D}}=\omega^2$ assumed above.

\subsection{$z=1$}

Firstly, we consider the terms up to $z=1$, namely
\begin{equation}
\mathcal{F}(\Delta) = c_{\phi}^2 \Delta - \frac{m^2}{\hbar^2}\,.
\end{equation}
As in Eq.~\eqref{eqn:formula_z=1}, we obtain
\begin{equation}
(\tilde{\Delta} f^+)g^- + f^+(\tilde{\Delta} g^-) \simeq \left( 2\gamma^{ij}\frac{\partial}{\partial r^i}\frac{\partial}{\partial r^j} + \frac{1}{2}\tilde{\Delta} \right)(f^+g^-)\,.
\end{equation}
Since the first term on the right-hand side is leading in $\hbar$, the corresponding operator $\tilde{\mathcal{D}}$ \eqref{eqn:def-calDtilde} is
\begin{equation}
\tilde{\mathcal{D}}=c_{\phi}^2 p^2 + m^2\,.
\end{equation}
Therefore the relation $\tilde{\mathcal{D}}=\omega^2$ holds in this example.

\subsection{$z=2$, spatially flat}

Next we consider terms up to $z=2$, Eq.~\eqref{eq:z=2dispersion}, and let us assume spatially flat geometries as in Sec.~\ref{sec:z=3flat}. Computing several relevant terms on the right-hand side of Eq.~
\eqref{eq:ansatzSymmetric} for $z=2$,
\begin{equation}
\begin{pmatrix}
16 \gamma^{ij}\gamma^{kl} \frac{\partial}{\partial r^i} \frac{\partial}{\partial r^j} \frac{\partial}{\partial r^k} \frac{\partial}{\partial r^l} (f^+g^-) \\
4 \gamma^{ij}\gamma^{kl}\partial_i \partial_j \frac{\partial}{\partial r^k} \frac{\partial}{\partial r^l} (f^+g^-) \\
4 \gamma^{ik}\gamma^{jl}\partial_i \partial_j \frac{\partial}{\partial r^k} \frac{\partial}{\partial r^l} (f^+g^-) \\
\gamma^{ij}\gamma^{kl} \partial_i \partial_j \partial_k \partial_l (f^+g^-)
\end{pmatrix}
=
\begin{pmatrix}
1 & -4 & 2 & 4 \\
1 & 0 & 2 & -4 \\
1 & 0 & -2 & 0 \\
1 & -4 & 2 & 4
\end{pmatrix}
\begin{pmatrix}
(\Delta^2 f^+)g^- + f^+(\Delta^2 g^-)\\
(\Delta\partial_if^+)(\partial^i g^-) + (\partial^i f^+)(\Delta\partial_i g^-) \\
(\Delta f^+)(\Delta g^-) \\
(\partial^i\partial^j f^+)(\partial_i\partial_j g^-)
\end{pmatrix}\,.
\end{equation}
Inverting the coefficient matrix, we then obtain
\begin{align}
(\Delta^2 f^+)g^- + f^+(\Delta^2 g^-)=&\biggl[ 2\gamma^{ij}\gamma^{kl} \frac{\partial}{\partial r^i} \frac{\partial}{\partial r^j} \frac{\partial}{\partial r^k} \frac{\partial}{\partial r^l} +(\gamma^{ij}\gamma^{kl}+2\gamma^{ik}\gamma^{jl})\partial_i \partial_j \frac{\partial}{\partial r^k} \frac{\partial}{\partial r^l} + \frac{1}{8}\gamma^{ij}\gamma^{kl} \partial_i \partial_j \partial_k \partial_l  \biggr] (f^+g^-)
\end{align}
Again, since the first term on the right-hand side is leading in $\hbar$, we have 
\begin{equation}
\tilde{\mathcal{D}}=\frac{\beta}{M^2}p^4 + c_{\phi}^2 p^2 + m^2\,,
\end{equation}
which agrees with $\omega^2$ as in Eq.~\eqref{eqn:omega_z=2}.

\subsection{$z=3$, spatially flat}

As the last example, let us consider terms up to $z=3$ \eqref{eqn:F(Delta)_z=3} with spatially flat geometries. By computing terms on the right-hand side of Eq.~
\eqref{eq:ansatzSymmetric} for $z=3$, it can be shown that
\begin{equation}
\begin{pmatrix}
64 \gamma^{ij}\gamma^{kl}\gamma^{mn} \frac{\partial}{\partial r^i} \frac{\partial}{\partial r^j} \frac{\partial}{\partial r^k} \frac{\partial}{\partial r^l} \frac{\partial}{\partial r^m} \frac{\partial}{\partial r^n} (f^+g^-) \\
16 \gamma^{ij}\gamma^{kl}\gamma^{mn}\partial_i \partial_j \frac{\partial}{\partial r^k} \frac{\partial}{\partial r^l} \frac{\partial}{\partial r^m} \frac{\partial}{\partial r^n} (f^+g^-) \\
16 \gamma^{ik}\gamma^{jl}\gamma^{mn}\partial_i \partial_j \frac{\partial}{\partial r^k} \frac{\partial}{\partial r^l} \frac{\partial}{\partial r^m} \frac{\partial}{\partial r^n} (f^+g^-) \\
4 \gamma^{ij}\gamma^{kl}\gamma^{mn}\partial_i \partial_j \partial_k \partial_l \frac{\partial}{\partial r^m} \frac{\partial}{\partial r^n} (f^+g^-) \\
4 \gamma^{ij}\gamma^{km}\gamma^{ln}\partial_i \partial_j \partial_k \partial_l \frac{\partial}{\partial r^m} \frac{\partial}{\partial r^n} (f^+g^-) \\
\gamma^{ij}\gamma^{km}\gamma^{ln}\partial_i \partial_j \partial_k \partial_l \partial_m \partial_n (f^+g^-)
\end{pmatrix}
=
\begin{pmatrix}
1 & -6 & 3 & 12 & -12 & -8 \\
1 & -2 & 3 & -4 & -4 & 8 \\
1 & -2 & -1 & 0 & 4 & 0 \\
1 & 2 & 3 & -4 & 4 & -8 \\
1 & 2 & -1 & 0 & -4 & 0 \\
1 & 6 & 3 & 12 & 12 & 8
\end{pmatrix}
\begin{pmatrix}
(\Delta^3 f^+)g^- + f^+(\Delta^3 g^-)\\
(\Delta^2\partial_if^+)(\partial^i g^-) + (\partial^i f^+)(\Delta^2\partial_i g^-) \\
(\Delta^2 f^+)(\Delta g^-) + (\Delta f^+)(\Delta^2 g^-) \\
(\Delta\partial^i\partial^j f^+)(\partial_i\partial_j g^-) + (\partial^i\partial^j f^+)(\Delta\partial_i\partial_j g^-) \\
(\Delta\partial^i f^+)(\Delta\partial_i g^-) \\
(\partial_i \partial_j \partial_k f^+)(\partial^i \partial^j \partial^k g^-)
\end{pmatrix}\,.
\end{equation}
Inverting the coefficient matrix, we then obtain
\begin{align}
(\Delta^2 f^+)g^- + f^+(\Delta^2 g^-)=&\biggl[ 2\gamma^{ij}\gamma^{kl}\gamma^{mn} \frac{\partial}{\partial r^i} \frac{\partial}{\partial r^j} \frac{\partial}{\partial r^k} \frac{\partial}{\partial r^l} \frac{\partial}{\partial r^m} \frac{\partial}{\partial r^n} + \frac{1}{2}(3\gamma^{ij}\gamma^{kl}\gamma^{mn}+12\gamma^{ik}\gamma^{jl}\gamma^{mn})\partial_i \partial_j \frac{\partial}{\partial r^k} \frac{\partial}{\partial r^l} \frac{\partial}{\partial r^m} \frac{\partial}{\partial r^n} \nonumber \\
& + \frac{1}{8}(3\gamma^{ij}\gamma^{kl}\gamma^{mn}+12\gamma^{ij}\gamma^{km}\gamma^{ln}) \partial_i \partial_j \partial_k \partial_l \frac{\partial}{\partial r^m} \frac{\partial}{\partial r^n} + \frac{1}{32} \gamma^{ij}\gamma^{kl}\gamma^{mn} \partial_i \partial_j \partial_k \partial_l \partial_m \partial_n \biggr] (f^+g^-)
\end{align}
Again, since the first term on the right-hand side is leading in $\hbar$, we have 
\begin{equation}
\tilde{\mathcal{D}}=\frac{\alpha}{M^4}p^6 + \frac{\beta}{M^2}p^4 + c_{\phi}^2 p^2 + m^2\,.
\end{equation}
Therefore, referring to Eq.~\eqref{eqn:omega_z=3}, the relation $\tilde{\mathcal{D}}=\omega^2$ holds also in this example.

\section{Energy, momentum, stress tensors for Lifshitz scalar}
\label{sec:energy-momentm}

The distribution function appears in field equations in expectation values of the energy-momentum tensor for Lorentz invariant matter fields. For the Ho\v{r}ava-Lifshitz setup, field equations are obtained by taking the variation of action with respect to fundamental variables $N$, $N^i$, and $\gamma_{ij}$. In this appendix, we derive how Wigner functions appear in the variation of action \eqref{eq:LifshitzAction} and show that in the leading order in $\hbar$, they are given only by $f_1$ as far as $|f_2| \ll |f_1|$. 

First, the variation with respect to $N^i$ is given by
\begin{align}
\frac{1}{\sqrt{\gamma}}\left\langle:\frac{\delta I}{\delta N^i}:\right\rangle&= - \frac{1}{2}\left\langle: \phi_\perp\partial_i\phi+\partial_i\phi\, \phi_\perp :\right\rangle \nonumber \\
&=\int \frac{d^3p}{(2\pi\hbar)^3\sqrt{\gamma}} p_i f_1^- - \frac{1}{4}\partial_i \left\langle: \phi_\perp\phi+\phi\phi_\perp :\right\rangle \nonumber \\
&=\int \frac{d^3p}{(2\pi\hbar)^3\sqrt{\gamma}} \left( p_i f_1 + \mathcal{O}(\hbar)f_3 \right),
\label{eq:dIdNi}
\end{align}
where the fact that $f_1^+$ is an even function of $p_i$ is used at the last equality. This is independent of the form of the function $\mathcal{F}(\Delta)$.

Second, the variation with respect to $N$ is given by
\begin{align}
-\frac{1}{\sqrt{\gamma}}\left\langle:\frac{\delta I}{\delta N}:\right\rangle&=\frac{1}{2}\left\langle:\phi_\perp^2-\phi\mathcal{F}(\Delta)\phi:\right\rangle \nonumber \\
&=\frac{1}{2}\int \frac{d^3p}{(2\pi\hbar)^3} \int d^3r\, e^{-\frac{i}{\hbar}r^ip_i} \left( \left\langle:\phi_\perp^+\phi_\perp^-:\right\rangle - \mathcal{F}(\Delta)_*^- \left\langle:\phi^+\phi^-:\right\rangle  \right).
\end{align}
Combining Eqs.~\eqref{eqn:integratedformula_general} and \eqref{eqn:integratedformula_general_symmetric},
\begin{align}
-\frac{1}{\sqrt{\gamma}}\left\langle:\frac{\delta I}{\delta N}:\right\rangle&=\frac{1}{2}\int \frac{d^3p}{(2\pi\hbar)^3\sqrt{\gamma}}\left[ F_{\phi_\perp\phi_\perp} + \frac{\tilde{\mathcal{D}}}{\hbar^2}\left(1+\mathcal{O}(\hbar) \right) F_{\phi\phi} \right] \nonumber \\
&=\int \frac{d^3p}{(2\pi\hbar)^3\sqrt{\gamma}} \left( \omega f_1 +\mathcal{O}(\hbar)(f_1,f_2)\right).
\label{eq:dIdN}
\end{align}
At the last equality, the fact that $f_1^-$ is an odd function of $p_i$ is used and $\tilde{\mathcal{D}}=\omega^2$ is assumed, which holds for examples considered in the last section. The form of Eq.~\eqref{eq:dIdN} is the same as the Lorentz invariant case, but the information of the dispersion relation is encoded in $\omega$.

Finally, the variation with respect to $\gamma_{ij}$ is evaluated by
\begin{align}
\frac{1}{N\sqrt{\gamma}}\left\langle:\frac{\delta I}{\delta \gamma_{ij}}:\right\rangle&=\frac{\gamma^{ij}}{4}\left\langle:\phi_\perp^2:\right\rangle + \frac{1}{2N\sqrt{\gamma}} \frac{\delta}{\delta \gamma_{ij}} \int dt d^3x\, N\sqrt{\gamma} \phi \mathcal{F}(\Delta) \phi \nonumber \\
&=\frac{\gamma^{ij}}{4} \int \frac{d^3p}{(2\pi\hbar)^3\sqrt{\gamma}} F_{\phi_\perp\phi_\perp} + \frac{1}{2N\sqrt{\gamma}} \frac{\delta}{\delta \gamma_{ij}} \int dt d^3x\, N \int \frac{d^3p}{(2\pi\hbar)^3} \frac{-\tilde{\mathcal{D}}}{\hbar^2}(1+\mathcal{O}(\hbar))F_{\phi\phi}.
\end{align}
Assuming $\tilde{\mathcal{D}}=\omega^2$ again and noting $\partial \phi^\pm/\partial \gamma_{ij} =[\partial/\partial \gamma_{ij}, e^{\pm \frac{r^i}{2}\tilde{\nabla}_i} ]\phi = \mathcal{O}(r) \phi^\pm$, we obtain
\begin{align}
\frac{1}{N\sqrt{\gamma}}\left\langle:\frac{\delta I}{\delta \gamma_{ij}}:\right\rangle&=\int \frac{d^3p}{(2\pi\hbar)^3\sqrt{\gamma}} \biggl[ \frac{\gamma^{ij}}{4}F_{\phi_\perp\phi_\perp} - \frac{1}{2\hbar^2} \left( \frac{\omega^2}{2}\gamma^{ij} + \frac{\partial \omega^2}{\partial \gamma_{ij}} \right) (1+\mathcal{O}(\hbar)) F_{\phi\phi} \biggr] \nonumber \\
&=\int \frac{d^3p}{(2\pi\hbar)^3\sqrt{\gamma}} \frac{1}{2} \biggl[ v_{\rm g}^{(i} p^{j)}f_1 + \left( v_{\rm g}^{(i} p^{j)} - \omega \gamma^{ij} \right) f_2 + \mathcal{O}(\hbar)f_i \biggr],
\label{eq:dIdgammaij}
\end{align}
where $\partial \omega^2/\partial \gamma_{ij}=-\omega v_{\rm g}^{(i} p^{j)}$ is used and $v_{\rm g}^i$ is defined in Eq.~\eqref{eq:groupvelocity}. One can easily see that this reproduces Eq.~\eqref{eq:Sij} for $z=1$ by noting that $v_{\rm g}^i=c_\phi^2 p^i/\omega$ and that the definition of $T^{\mu\nu}$ \eqref{eq:Tmunu} includes the factor of 2. Therefore, we should keep $f_2$ small, since Eq.~\eqref{eq:dIdgammaij} includes $f_2$ as well as $f_1$ at the leading order of $\hbar$ as in the case with $z=1$.

\bibliographystyle{apsrmp}

\end{document}